\documentclass[aps,prl,reprint,showpacs,superscriptaddress,longbibliography]{revtex4-1}

\usepackage{graphicx}
\usepackage{physics}

\usepackage{amsmath,amssymb}
\usepackage{color}

\usepackage{hyperref}
\hypersetup{
setpagesize=false,
colorlinks=true,
linkcolor=blue,
citecolor=red,
}
\usepackage{bm}

\begin{document}

\title{Directional propagation of quantum Hall viscous fluid by nano-structural engineering}

\author{Hiroshi Funaki}
\affiliation{%
Kavli Institute for Theoretical Sciences, University of Chinese Academy of Sciences, Beijing, 100190, China.
}%
\author{Ai Yamakage}
\affiliation{
Department of Physics, Nagoya University, Nagoya 464-8602, Japan
	}
\author{Ryotaro Sano}
\affiliation{%
Department of Physics, Kyoto University, Kyoto 606-8502, Japan
}%
\author{Mamoru Matsuo }
\email{mamoru@ucas.ac.cn}
\affiliation{%
Kavli Institute for Theoretical Sciences, University of Chinese Academy of Sciences, Beijing, 100190, China.
}%

\affiliation{%
CAS Center for Excellence in Topological Quantum Computation, University of Chinese Academy of Sciences, Beijing 100190, China
}%
\affiliation{%
RIKEN Center for Emergent Matter Science (CEMS), Wako, Saitama 351-0198, Japan
}%
\affiliation{%
Advanced Science Research Center, Japan Atomic Energy Agency, Tokai, 319-1195, Japan
}%

\date{\today}

\begin{abstract}
We present a microscopic theory of the viscous electron fluid in the quantum Hall state based on the nonequilibrium Green's function method and the von Neumann lattice representation.
This approach permits the formulation of hydrodynamic equations in the strong field regime that accommodates arbitrary boundary conditions.
We demonstrate nonreciprocal transport resulting from the interplay between magnetic field-induced viscosity and device geometry in a notched system. 
Our results will offer a powerful tool for studying the nonperturbative effects of magnetic fields on electron viscous fluids. 
\end{abstract}

\maketitle 

{\it Introduction.---}
The viscous electron fluid, realized in high-purity solid-state systems, has consistently revealed a wide range of diverse nonequilibrium properties and has been the subject of vigorous research. 
For electronic transport to exhibit viscous fluid behavior, the Coulomb interactions between electrons must dominate over impurity scattering, which necessitates the use of high-purity many-body electronic systems.
Initially proposed theoretically and later demonstrated experimentally in GaAs quantum wells with two-dimensional electron gases (2DEGs)~\cite{Keser2021PRX, gurzhi1963,gurzhi1968,gurzhi1968magnetic-f,gurzhi1995,Buhmann2002,melnikov2012,molenkamp1994,jong1995,molyneux1995,hara2004,tomadin2014,braem2018,ginzburg2021,vijayakrishnan2023,alekseev2020,ahn2022,gusev2018AIP,levin2018,gusev2018PRB,gusev2020,raichev2020,gold2021,keser2021,wang2023,levin2023,patricio2024}, this phenomenon has also more recently been actively investigated in two-dimensional materials such as graphene~\cite{Bandurin2016Science,Crossno2016Sicence,Levitov2016NatPhys,KrishnaKumar2017NatPhys,Bandurin2018NatCommun,Sulpizio2019Nature,Gallagher2019Science,Ella2019NatNanotech,Berdyugin2019Science,Ku2020Nature} and 2D metal PdCoO$_2$~\cite{Moll2016Science}.

Recently, research has explored the non-trivial effects of magnetic fields on the electron viscous fluid in two-dimensional electron gases~\cite{alekseev2016,wang2022,ginzburg2023,cheremisin2024,scaffidi2017,pellegrino2017,burmistrov2019,holder2019classical,matthaiakakis2020,berdyugin2019,gusev2018AIP,levin2018,gusev2018PRB,gusev2020,raichev2020,gold2021,keser2021,wang2023,levin2023,
patricio2024}. Historically, 2DEGs in magnetic fields have been key subjects of study in the context of the quantum Hall effect, remaining a central focus in the investigation of electronic transport in strongly correlated systems. 
This interest extends to understanding how the electron viscous fluid behaves within the quantum Hall state, where it is expected to significantly broaden the scope of research into electron transport phenomena under strong magnetic fields.

\begin{figure}[h]
\includegraphics[width=.9\hsize]{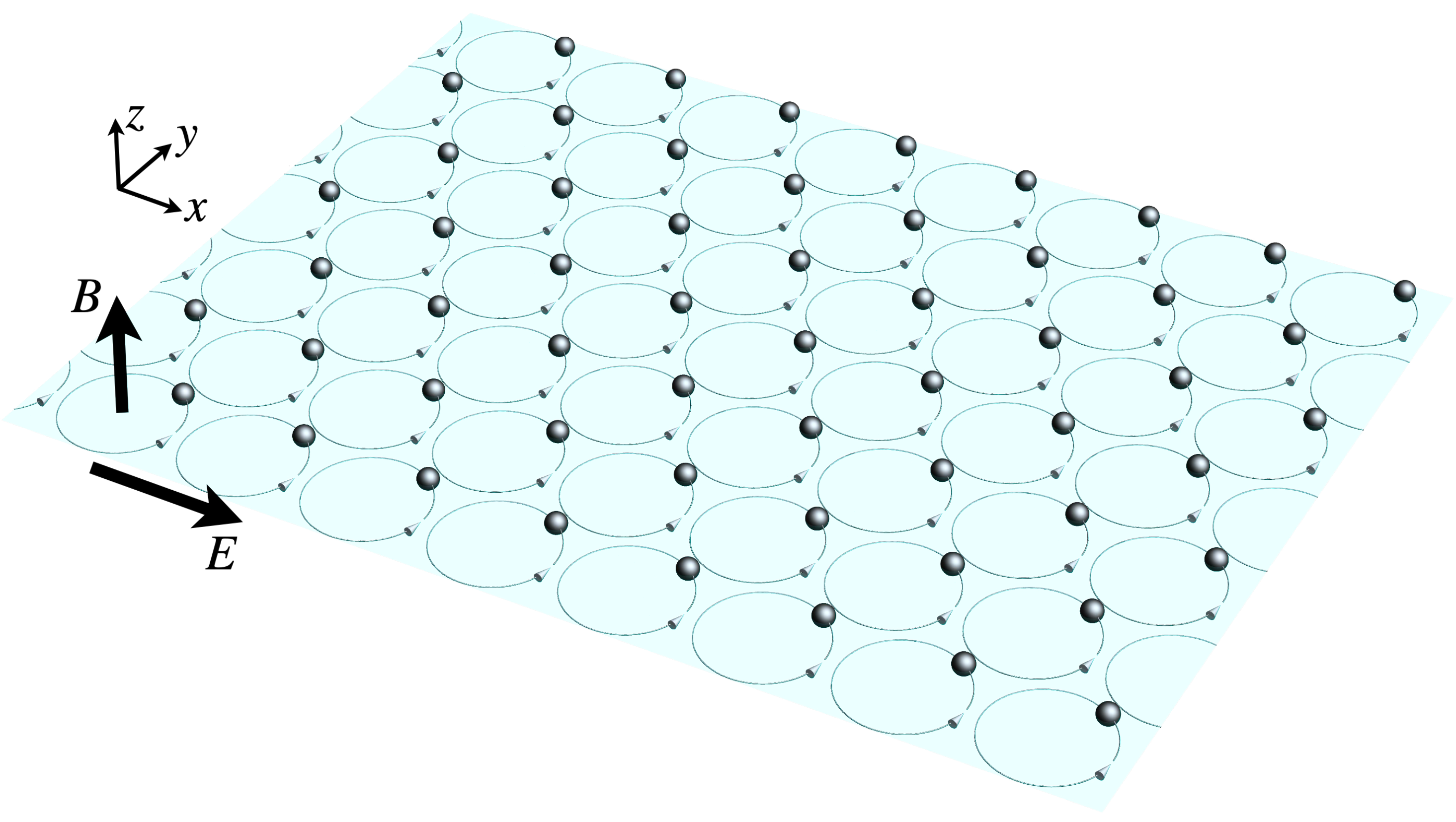}
 \caption{Schematics of the von Neumann lattice.
  }
\label{schema}
\end{figure}

In this Letter, we construct a microscopic theory of the viscous electron fluid in the quantum Hall state using the nonequilibrium Green's function method. Specifically, we employ the von Neumann lattice representation, which allows for a systematic description of quantum Hall dynamics as a subset of coherent states (see Fig.~\ref{schema}). Using the magnetic length as the characteristic scale, we introduce appropriate approximations to evaluate the microscopic momentum flux in the regime where electrons undergoing cyclotron motion in a strong magnetic field behave as a viscous fluid. This enables the derivation of fluid equations that do not depend on specific boundary conditions. While the Landau gauge and symmetric gauge provide clear descriptions for certain geometries in quantum Hall systems, this approach is well-suited for investigating fluid-like behavior under various boundary conditions and inhomogeneous transport across different device geometries.
The derived fluid equations feature a viscous term originating from the magnetic field, and we demonstrate that this naturally extends the well-known Stokes equation into the strong field regime. Furthermore, we numerically show that the interplay between magnetic field-induced viscosity and device geometry exhibits the striking feature of non-reciprocal transport in systems with a notch.
Our results will offer a powerful tool for investigating the non-perturbative effects of magnetic fields on viscous electron fluids and serve as a foundation for further expanding the research landscape of electron viscous fluids.

{\it Model and the von Neumann lattice.---}
We start from the following model Hamiltonian to derive a hydrodynamic equation for two-dimensional electron systems in a strong magnetic field,
\begin{subequations}
\begin{align}
H &= H_{\rm LL} +H_{\rm ex}
.
\end{align}
The kinetic energy
is given by
\begin{align}
H_{\rm LL} = 
  \int \!\! d^2{\bm r} \ 
  c^\dagger_{\bm r} \frac{P_x^2 +P_y^2}{2m}  c_{\bm r}
,
\end{align}
where dynamical momentum is 
${\bm P} = {\bm p} +e{\bm A}$, 
a uniform constant magnetic field is 
$\nabla \times {\bm A} = B {\bm e}_z$,
and $-e$ and $m$ are the charge and mass of the electron, respectively.
The Hamiltonian $H_\mathrm{ex}$ drives the electron dynamics under the external electromagnetic field:
\begin{align}
H_{\rm ex} &= 
  \frac{1}{m} \int \!\! d^2{\bm r} \ 
  e {\bm A}_{\rm ex} ({\bm r}, t) \cdot 
  c^\dagger_{\bm r} {\bm P} c_{\bm r}
,
\end{align}
\end{subequations}
where
${\bm A}_{\rm ex} ({\bm r},t) \approx 
  - t ( {\bm E}_{\rm ex}(\bm 0) 
    +x \partial_x {\bm E}_{\rm ex}(\bm 0) 
    +y \partial_y {\bm E}_{\rm ex}(\bm 0))$. 
 Here, $E_\mathrm{ex}$ denotes the external electric field.
It should be noted that we will treat the electron-electron scattering and electron-impurity scattering as semi-phenomenological electron lifetimes later, so that the microscopic Hamiltonian for the scattering processes is not explicitly given above.

The kinetic term $H_\mathrm{LL}$ is quantized into the Landau levels, reflecting the cyclotron motion of the electrons under a strong magnetic field. In particular, we utilize the von Neumann lattice representation~\cite{ezawa2013_book}, aiming to derive the electron viscous fluid equations for arbitrary geometries (see Fig.~\ref{schema}). In this representation, $H_\mathrm{LL}$ reads 
\begin{align}
H_{\rm LL} &=\sum_{N, \beta} c_{N, \beta}^\dagger
  \hbar \omega_c \Big( N +\frac{1}{2} \Big) c_{N, \beta}
,
\end{align}
where $\omega_c=\frac{e\abs{B}}{m}$ is the cyclotron frequency, $N$ is the label of the energy level, and $\beta$ is the label of the degree of freedom within the energy level, representing a position on the von Neumann lattice.
The annihilation operator of the electron is introduced as
$c_{\bm r} = \sum_{N,\beta} \phi_{N,\beta}({\bm r}) c_{N,\beta}$,
where $\phi_{N,\beta}$ is the wave function of this state 
given by
$\phi_{N,\beta}({\bm r}) = \frac{1}{\sqrt{N!}} (a^\dagger)^N \phi_{0,\beta}({\bm r})$,
$\phi_{0,\beta}({\bm r}) = 
\frac{1}{\sqrt{2 \pi \ell_B^2}} 
  {\rm exp} [
   -|\frac{r}{2 \ell_B} -\frac{\beta}{\sqrt{2}} |^2
   +\frac{i}{\sqrt{2} \ell_B} (y {\rm Re}[\beta] +x {\rm Im}[\beta]) ]$
with the magnetic length
$\ell_B=\sqrt{\frac{\hbar}{e \abs{B} }}$
and the ladder operator is defined by
$a = \frac{P_x -i P_y}{\sqrt{2\hbar e\abs{B}}}$, satisfying the commutation relation $[a,a^\dagger] =1$
~\footnote{See the supplemental material for more details on the von Neumann representation.}.

{\it Hydrodynamic equation.---} 
Hydrodynamic equation is derived from the expectation value of the operator relation of the momentum conservation law
\begin{subequations}
\begin{align}
\frac{\partial {\hat {\bm j}}}{\partial t}&=
  -\frac{i}{\hbar}[{\hat {\bm j}},H]
= -\nabla {\hat \Pi} +{\hat {\bm F}} 
,
\end{align}
with
\begin{align}
{\hat {\bm j}} &= c_{\bm r}^\dagger {\bm P} c_{\bm r}
, \ 
{\hat \Pi}_{ij} = c_{\bm r}^\dagger 
  \frac{ P_i P_j +P_j P_i }{2m}
  c_{\bm r}
,
\\
{\hat {\bm F}} &=
 e \frac{i \hbar}{2m} (
  {\bm B} \times c_{\bm r}^\dagger \overleftrightarrow{\nabla} c_{\bm r} )
,
\end{align}
\end{subequations}
where ${\hat {\bm j}}$, ${\hat \Pi}_{ij}$, and $\hat{\bm{F}}$ represent the momentum density, the momentum flux, and the force operators, respectively. Here, we have defined  $(\nabla \hat{\Pi})_i = \nabla_j \hat{\Pi}_{ji} $ and $c^\dagger \overleftrightarrow{\nabla} c = c^\dagger (\nabla c) - (\nabla c^\dagger) c$.

To evaluate the expectation values, let us consider coarse-grained physical quantities on scales larger than the magnetic length $\ell_B$ to derive the hydrodynamic equation. Specifically, we neglect the overlap between any two Gaussian-type wavefunctions under a strong magnetic field except when the same level at the same lattice point because the wavefunctions decay on a scale of $\ell_B$ around each lattice point $\beta$:
\begin{align}
  \int_\Omega d^2 {\bm r} \phi_{N',\beta'}({\bm r}) \phi_{N,\beta}(\bm r)
  \approx 
  \delta_{N' N} \delta_{\beta' \beta} 
.
\end{align}
where $\Omega$ denotes the area of the coarse-graining region.
Together with the approximation, we calculate the expectation value of the momentum flux as a linear response of $H_{\rm ex}$, namely, $H_{\rm ex}$ is treated as the perturbed Hamiltonian.

The effects of electron-electron scattering and impurity scattering are taken into account by the electron lifetime $\tau$ as semi-phenomenological parameter, assuming the situation under $\omega_c \tau \gg 1$.
Furthermore, spatially coarse-grained expectations are calculated under the assumption that the spatial dependence of the electric field is taken into account up to the first order.
Under these considerations, we obtain the coarse-grained expectation value of the momentum flux 
$\langle \hat{\Pi}_{ij} \rangle_{\Omega} = {\Pi}_{ij}$
~\footnote{See the supplemental material for details on the coarse-graining procedures.},
\begin{subequations}
\begin{align}
{\Pi}_{ij}
&= \Pi^{a}_{ij} +\Pi^{b}_{ij}
,
\\
 \Pi^{a}_{ij} &= - \eta^{a}_{ij kl} \partial_k E_{\mathrm{ex},l}
 +E_{\rm LL} \delta_{ij}
,
\\
 \Pi^{b}_{ij} &= - \eta^{b}_{ij kl} \partial_k E_{\mathrm{ex},l}
,
\end{align}
\end{subequations}
where
$E_{\rm LL}=\langle H_{\rm LL}\rangle_\Omega$
is the expectation value of the kinetic energy,
$\eta^{a}_{ij kl} = \eta^{a} (-1)^{\bar{\delta}_{kx}} \delta_{ij} \bar{\delta}_{kl}$,
$ \eta^{b}_{ij kl} = 
  \eta^{b}_1 [
    (-1)^{\bar{\delta}_{ik}} \delta_{ij} \delta_{kl}
   + \bar{\delta}_{ij} \bar{\delta}_{kl} ]
 +\eta^{b}_2 [
    (-1)^{\bar{\delta}_{ix}} \delta_{ij} \bar{\delta}_{kl}
   +(-1)^{\bar{\delta}_{ky}} \bar{\delta_{ij}} \delta_{kl} ] $,
with the Kronecker delta $\delta_{ij}$
and $\bar{\delta}_{ij}=1-\delta_{ij}$,
and the coefficients are derived as
\begin{subequations}
\begin{align}
\eta^{a} &=
  -\frac{4 \hbar }{m \omega_c} D_L \sum_{N} 
  \int \!\! d\omega
  f'_\omega E_N^2 ({\rm Im} G_N^{\rm A})^2
,
\\
\eta^{b}_1 &= 
 -\frac{\hbar^2}{2m}
  \sum_{N} \int \!\! d\omega 
 (-f'_{\omega})
  C_N
  D_L{\rm Im}G_{N}^{\rm A}
,
\\
\eta^{b}_2 &=
 \frac{\eta^{b}_1}{\omega_c \tau}
 - \frac{\hbar}{m \omega_c}
 \sum_{N} \int \!\! d\omega 
  f_{\omega}\frac{E_N}{(\hbar \omega_c)^2}
  D_L{\rm Im}G_{N}^{\rm A}
.
\end{align}
\end{subequations}
Here, 
$G^{\rm A}_{N}(\omega) = [\hbar \omega -E_N -i\hbar /(2\tau)]^{-1}$
is the advanced Green function,
$E_N = \hbar \omega_c ( N+\frac{1}{2} )$
is the energy level,
$f_\omega$ is the Fermi distribution function for the energy $\hbar \omega$,
$D_L=\frac{\Omega}{2 \pi \ell_B^2}$ 
is the number of states per Landau level in the coarse-grained region, and 
$C_N=\frac{\omega_c \tau}{1+ (2 \omega_c \tau)^2}
( \frac{2}{(\hbar \omega_c)^2} E_N^2 +\frac{3}{2} )$.
Similarly, the expectation value of the momentum density is also obtained as a linear response to the electric field, 
the momentum flux can be described in terms of the momentum density by replacing the electric field with the expectation value of momentum density.

Finally, we arrive at
the hydrodynamic equation for determining the momentum density distribution ${\bm j}=\langle \hat{{\bm j}} \rangle_{\Omega}$
\begin{subequations}
\begin{align}
\frac{\partial {\bm j}}{\partial t}
& \! = \! 
  \eta \nabla^2 {\bm j}
  \! + \! \eta_\perp \! \nabla^2 {\bm j}_\perp
  \! + \! \zeta_\perp \! \hat{\nabla}^2 {\bm j}_\perp
  \! - \! \nabla P
  \! + \! \mathfrak{s}(B) \omega_c {\bm j}_\perp 
,
\label{Eq_hydro}
\end{align}
where ${\bm j}_{\perp}$  is the vector obtained by rotating  ${\bm j}$  by 90 degrees around the $z$-axis as ${\bm j}_\perp = {\bm j} \times {\bm e}_z$, 
$\hat{\nabla}^2 {\bm j}_\perp = \nabla (\nabla \cdot {\bm j}_\perp)$, and $\mathfrak{s}(B)=B/|B|$.
Here, we have used the incompressible condition $\nabla \cdot {\bm j}=0$ and the equation of state $P=  E_{\rm LL}$.
The viscosity coefficients are given by
\begin{align}
\eta \! = \!
  \eta^{b}_1 \Sigma_\perp \! + \! \eta^{b}_2 \Sigma_\parallel
, \ 
\eta_{\perp} \! = \!
 \eta^{b}_1 \Sigma_\parallel \! - \! \eta^{b}_2 \Sigma_\perp
, \ 
\zeta_{\perp} \! = \!
  \eta^{a} \Sigma_\parallel
,
\end{align}
with
$\Sigma_\parallel = \frac{\sigma_\parallel}{\sigma_\parallel^2 +\sigma_\perp^2}$,
$\Sigma_\perp =  \frac{\sigma_\perp}{\sigma_\parallel^2 +\sigma_\perp^2}$,
and the conductivities are given by
\begin{align}
\sigma_\parallel &= 
   \sum_{N} \int \!\! d\omega
  (-f'_{\omega}) E_N (2 D_L{\rm Im}G_N^{\rm A}) 
  \frac{\tau}{1+ (\omega_c \tau)^2}
,
\\
\sigma_\perp &= 
  \frac{\sigma_\parallel}{\omega_c \tau}
 -\frac{1}{\omega_c} \sum_{N} \int \!\! d\omega
   f_{\omega} (2 D_L{\rm Im}G_N^{\rm A})
.
\end{align}
\end{subequations}

{\it Flow and pressure profiles under strong magnetic field.---}
%
\begin{figure}[h]
\includegraphics[width=1.0\hsize]{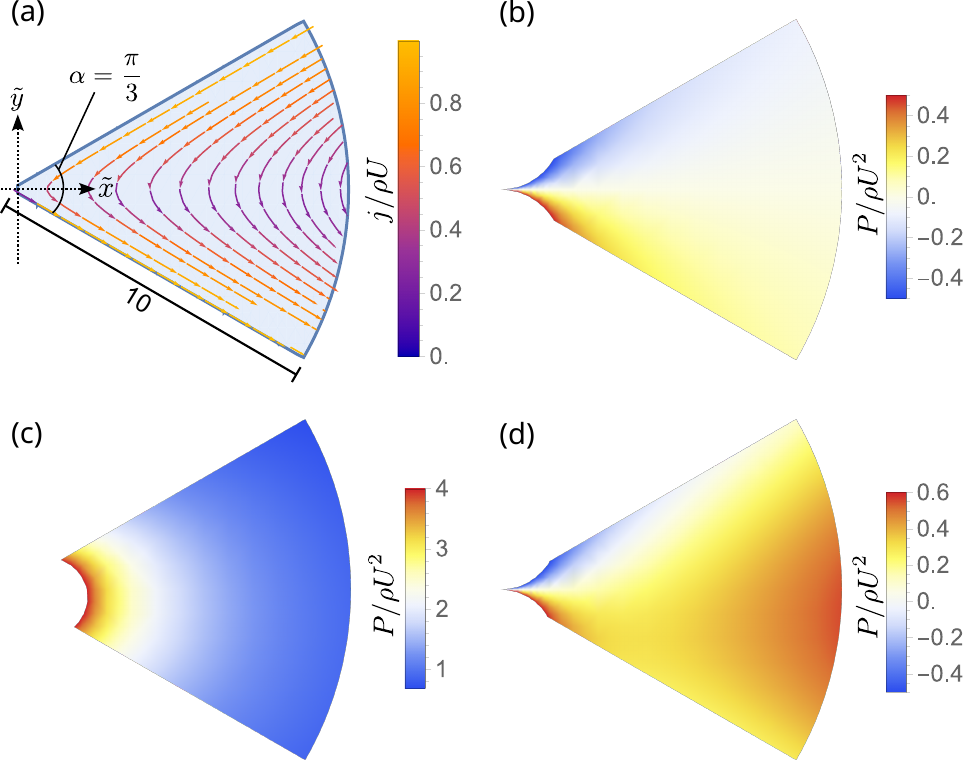}
 \caption{
Momentum density and pressure profiles in the wedge-shaped domain, where $\tilde{x} = x/L$, $\tilde{y} = y/L$, $L$ is the sample size, $\alpha=\frac{\pi}{3}$, and 
 the magnitude of the edge flow is $\rho U$.
The non-dimensionalized shear viscosity, Hall viscosity, and Lorentz force coefficients are
$\tilde{\eta}=\frac{\eta}{UL}$, 
$\tilde{\eta}_{\rm H}=\frac{ \eta_{\rm H} }{UL}$, and 
$\tilde{\omega}_c=\frac{\omega_c}{U/L}$, respectively.
(a) The momentum density streamline is independent of the magnetic field effects $\tilde{\eta}_{\rm H}$ and $\tilde{\omega}_c$.
On the other hand, pressure profile depends on them.
(b) Pressure profile when shear viscosity is dominant, i.e., conventional Stokes flow, at 
$\tilde{\eta}=1$,
$\tilde{\eta}_{\rm H}= \tilde{\omega}_c= 0$.
It is antisymmetric in the $y$ direction.
(c) Pressure profile when Hall viscosity is dominant at 
$\tilde{\eta}=1$,
$\tilde{\eta}_{\rm H}= 5$,
$\tilde{\omega}_c= 0$.
 It is symmetric in the $y$ direction, with high pressure at small $x$.
(d) Pressure profile when Lorentz force is dominant at 
$\tilde{\eta}=1$,
$\tilde{\eta}_{\rm H}= 0$,
$\tilde{\omega}_c= 0.2$.
At large $x$, it is symmetric in the $y$ direction and high pressure.
} 
\label{fig_Taylor}
\end{figure}
%
%
Let us consider analytical solutions of the hydrodynamic equations in which the magnetic field breaks the mirror symmetry, and it is shown that it is possible to determine whether shear viscosity, Hall viscosity, or Lorentz force dominates from the qualitative behavior of the pressure profile.

To simplify the derived two-dimensional hydrodynamic equation, 
we use the complex number representation:
$z = x +i y$, $\bar{z} = x -i y$, 
$j_z = j_x +i j_y$, $\bar{j}_z = j_x -i j_y$,
and 
the incompressible condition
${\rm Re}[\frac{\partial j_z}{\partial z}] =0$. Then the hydrodynamic equation~(\ref{Eq_hydro}) is cast into the complex form,
\begin{align}
\!\!\!\!\frac{\partial j_z}{\partial t} = 
 4 (i \eta +\eta_{\rm H}) 
 \frac{\partial}{\partial \bar{z}} {\rm Im}\Big[ \frac{\partial j_z}{\partial z} \Big]
 -2 \frac{\partial}{\partial \bar{z}} P
 - i\mathfrak{s}(B) \omega_c j_z,    
\end{align}
where
$\eta_{\rm H} = \eta_{\perp} +\zeta_{\perp}$. 
This equation can be interpreted as a two-dimensional Stokes equation that is modified by mirror symmetry breaking terms, such as the Hall viscosity $\eta_\mathrm{H}$ and the Lorentz force, $- i\mathfrak{s}(B) \omega_c j_z$. 

We find
a special solution for a steady state in a wedge-shaped domain 
in the presence of edge current:
\begin{align}
j_z &=
 -2 \gamma_1 \theta 
 +i \gamma_1 e^{2i \theta}
 -i \gamma_2
,\label{asol_jz}
\\
P &=
 4\gamma_1 \frac{ \eta_{\rm H} x -\eta y }{r^2}
 +\mathfrak{s}(B) \omega_c  [(\gamma_2-\gamma_1) x -2\gamma_1 y \theta]
,\label{asol_P}
\end{align}
where
$z=x+iy=re^{i \theta}$,
$ \gamma_1 = \frac{\cos (\alpha/2)}{\alpha +\sin \alpha} \rho U $, 
$ \gamma_2 = \frac{\alpha \sin (\alpha/2) +\cos (\alpha/2)}{\alpha +\sin \alpha} \rho U $. 
Here, we consider the boundary conditions where the upper and lower walls are arranged at an angle $\pm \frac{\alpha}{2}$, and the edge flow is
$j_z (\theta =\pm \frac{\alpha}{2}) = \mp \rho U e^{\pm i \frac{\alpha}{2}}$,
with the mass density $\rho$.
It is worth noting that the solution in Eq.~(\ref{asol_jz}) remains unaffected by mirror symmetry breaking terms, whereas the solution in Eq.~(\ref{asol_P}) explicitly depends on them.
Therefore,
the flow profile depicted in Fig.~\ref{fig_Taylor}(a) remains unchanged 
regardless of the magnitudes of the Hall viscosity and the Lorentz force. On the other hand, the pressure reflects the mirror symmetry breaking effects as plotted in Figs.~\ref{fig_Taylor}(b)--\ref{fig_Taylor}(d).
When the shear viscosity contribution is dominant [Fig.~\ref{fig_Taylor}(b)],
the pressure is antisymmetric in the $y$ direction.
In contrast, when the Hall viscosity contribution is dominant [Fig.~\ref{fig_Taylor}(c)], the pressure is almost symmetric in the $y$ direction, with higher pressure at small $x$.
Furthermore, when the Lorentz force contribution is dominant [Fig.~\ref{fig_Taylor}(d)], the pressure is almost symmetric in the $y$ direction, with higher pressure at large $x$, but the influence is negligible at small $x$.
In this way, the pressure distribution reflects information about mirror symmetry breaking effects, making it well-suited for observing the nontrivial impacts of a magnetic field on electron viscous fluids. In fact, since the momentum density and pressure of the electron viscous fluid are proportional to the current density and voltage, respectively, we propose that performing separate measurements of current and voltage can reveal these pronounced differences caused by nonperturbative magnetic effects.

%
{\it Nonreciprocal flow in a notched geometry.---}
To explore further properties associated with broken mirror symmetry, 
we numerically investigate the flow profile in a notched geometry owing to the ability to accommodate arbitrary geometric boundary conditions.

\begin{figure}[h]
 \includegraphics[width=1.0\hsize]{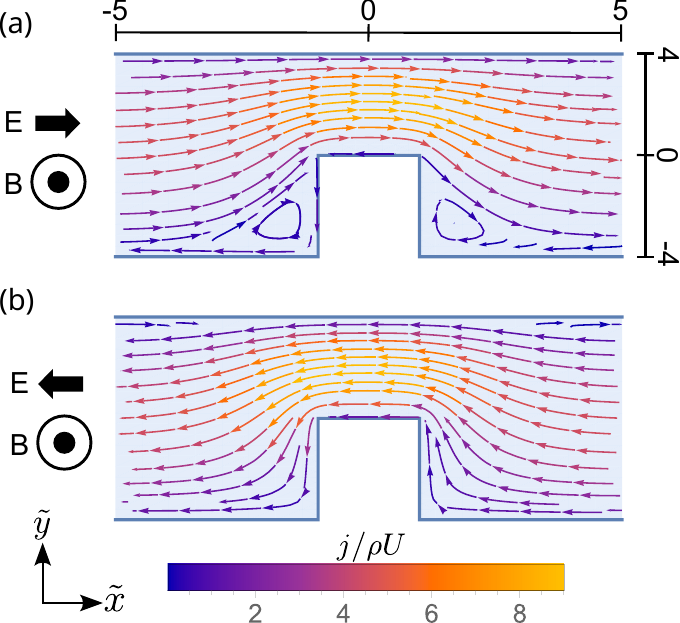}
 \caption{
 Momentum density streamlines in a notched channel with edge flow. Parameters are set to \(\tilde{\eta}=\tilde{\eta}_{\rm H}=\tilde{\omega}_c=1\). The edge flow is clockwise with a magnitude of \(\rho U\). Boundary conditions are imposed as follows: \(P=0\) at \(\tilde{x}=20\) and, at \(\tilde{x}=-20\), (a) \(j_x=+3\rho U\) or (b) \(j_x=-3\rho U\).
It exhibits a non-reciprocal behavior whether the vortices are generated or not in the vicinity of the notch, depending on the direction of the electric field.}
\label{Nonreciprocal}
\end{figure}

Figure~\ref{Nonreciprocal} shows numerical demonstrations in a notched geometry by changing the direction of the external electric field, while fixing the directions of the magnetic field and the edge flow.
Depending on the direction of the electric field, 
the flow with and without vortices is observed in a large region near the notch.
The vortices in panel (a) originate from the opposite orientation of the main flow and the edge flow in the vicinity of the notch. Consequently, the formation of vortices can be switched by external electric fields, thereby manifesting a nonreciprocal response. This behavior arises from the interplay between magnetic field-induced viscosity and device geometry, serving as a characteristic feature
of viscous electron fluids in a strong magnetic field.

{\it Discussion.---} 
%
Nonreciprocal transport is
characterized by directional propagation of quantum particles~\cite{Tokura2018NatCommun}. These responses occur in material systems without inversion and/or time-reversal symmetries. 
Previous studies on nonreciprocal responses have mainly focused on the geometric properties of the electron wave function such as the Berry curvature, which are introduced by intrinsic symmetry breaking in bulk materials~\cite{Tokura2018NatCommun}. 
By contrast, an inhomogeneous electron flow driven by the interplay between the viscous effects and the device geometries is a hallmark of electron hydrodynamics. 
Therefore, extrinsic symmetry breaking induced by device nanostructuring~\cite{geurs2020arXiv,Huang2024Nature} and magnetic fields~\cite{Sano2021PRBL} is an alternative strategy for nonreciprocal electron hydrodynamics.
A fluid flowing through a notched device will show very different current profile for opposite flow directions (see Fig.~\ref{Nonreciprocal}), which in turn results in nonreciprocal magnetoresistance.

Furthermore, we highlight the broader implications of our results. In particular,  
our formulation of electronic viscous fluid dynamics in strong magnetic fields not only builds on but also extends ongoing research on magnetoresistance in electronic viscous fluids~\cite{gurzhi1963,gurzhi1968,gurzhi1968magnetic-f,bockhorn2011,hatke2012,bockhorn2013,mani2013,shi2014,alekseev2016,wang2022,ginzburg2023,cheremisin2024,gusev2021,kumar2023,levin2024,levchenko2017,mandal2020,zeng2024,patricio2024}.
This theory offers new insights into the nature of electronic viscous fluid behavior under strong magnetic fields. It also sheds light on the ongoing vigorous efforts to understand the crossover from viscous fluid to ballistic conduction regimes~\cite{Chandra2019,holder2019channel,gupta2021,gupta2021b,Egorov2024}, and we anticipate that these findings will inform and inspire future research in this expanding field.

{\it Conclusion.---}
In this work, we have developed a microscopic theory of the viscous electron fluid in the quantum Hall state using the nonequilibrium Green's function method and the von Neumann lattice representation. Focusing on electrons undergoing cyclotron motion with the magnetic length as the characteristic scale for coarse-graining, we derived fluid equations that incorporate magnetic field-induced viscous terms, naturally extending the Stokes equation into the strong field regime. Our theoretical framework provides a versatile tool for investigating fluid-like behavior in quantum Hall systems under a variety of boundary conditions and device geometries.

Furthermore, our numerical simulations revealed the striking feature of non-reciprocal transport in systems with a notch, a phenomenon that further highlights the unique transport properties of the viscous electron fluid under strong magnetic fields. These findings open new avenues for studying non-perturbative effects of magnetic fields on electron transport in strongly correlated systems, broadening the understanding of viscous electron fluids.

{\it Acknowledgements.---}
The authors are grateful to Yukio Nozaki, Takahiro Uto, Yuya Ominato, and
Danyu Shu for their valuable comments.
This work was supported by JSPS KAKENHI for Grants (No.~20K03835, 22KJ1937, 23H01839, 24H00322, and 24H00853) from MEXT, Japan,
and by the National Natural Science Foundation of China (NSFC) under Grant No. 12374126, and 
by the Priority Program of Chinese Academy of Sciences under Grant No. XDB28000000. 

\bibliography{ref}

\end{document}